\documentclass[a4paper,11pt]{article}
\usepackage{pos}
\usepackage{gnuplot-lua-tikz}
\usepackage{tabularx,booktabs}
\usepackage{dsfont}

\title{Properties of the $\eta$ and $\eta^{\prime}$ mesons: Masses, decay constants and gluonic matrix elements
}
\ShortTitle{Properties of the $\eta$ and $\eta^{\prime}$ mesons}

\manuallySeparateAuthors
\author*[2]{Gunnar S. Bali}
\author{, Sara Collins and }
\author*[1]{Jakob Simeth}
\author{ for the RQCD collaboration\\[0.5cm]
  \hspace{3mm}\includegraphics[width=3cm]{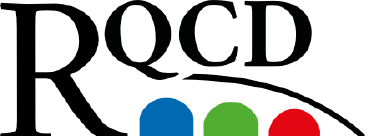}}

\note{Original contribution: ``Properties of the $\eta$ and $\eta^{\prime}$ mesons Part I:\ Masses and decay constants''}
\note{Original contribution: ``Properties of the $\eta$ and $\eta^{\prime}$ mesons Part II:\ Gluonic matrix elements''}
\affiliation{Institut für Theoretische Physik,\\
  Universität Regensburg, D-93040 Regensburg, Germany}

\emailAdd{gunnar.bali@ur.de}
\emailAdd{sara.collins@ur.de}
\emailAdd{jakob.simeth@ur.de}

\abstract{%
We present results for the $\eta$ and $\eta^\prime$ masses and their four independent decay constants at the physical point as well as their anomalous gluonic matrix elements $a_{\eta^{(\prime)}}$.
The chiral and continuum limit extrapolation is performed on twenty-one $N_f = 2+1$ Coordinated Lattice Simulations (CLS) ensembles with non-perturbatively
improved Wilson fermions at four different lattice spacings and along two trajectories in the quark mass plane, including one ensemble very close to physical
quark masses.
For the first time the decay constants are determined directly from the axialvector matrix elements without model assumptions. This allows us to study their QCD scale dependence and to determine all low-energy constants contributing at next-to-leading order in large-$N_c$ ChPT at a well defined QCD renormalization scale.
We also discuss higher excited states in the $1400\,\mathrm{MeV}$ region.
 }

\FullConference{%
 The 38th International Symposium on Lattice Field Theory, LATTICE2021
  26th-30th July, 2021
  Zoom/Gather@Massachusetts Institute of Technology
}

\begin{document}
\maketitle

\section{Introduction}
Within the SU(3) nonet of pseudoscalar mesons, the $\eta$ and $\eta^{\prime}$
particles play a special role. Resorting to a state mixing picture, based on
an effective Lagrangian, these can be viewed as mixtures between
octet and singlet components. While the former as a pseudo-Goldstone
boson of SU(3) flavour symmetry breaking is expected to be light,
the singlet component becomes
heavy due to the anomalous breaking of the axial U(1) symmetry. In
the chiral effective field theory (chiral perturbation theory, ChPT),
this fact can be taken into account by simultaneously expanding around
small quark masses and the large-$N_{c}$ limit where the anomaly
vanishes~\cite{DiVecchia:1980yfw,Kawarabayashi:1980dp,DiVecchia:1980vpx,Leutwyler:1997yr}. This approach, in combination with experimental input, often at
an unknown, low scale, enables the prediction of many properties like the
decay constants, see, e.g.~\cite{Feldmann:1998vh,Bickert:2016fgy}. 
However, using such input necessitates one to
neglect the QCD-scale dependence of some of the low-energy constants
(LECs). This can be motivated in part by the Okubo-Zweig-Iizuka (OZI)
suppression of scale-dependent contributions~\cite{Feldmann:1998vh}.
The extent of validity of this approximation and of the large-$N_c$ ChPT
expansion itself, however, has not been established from
first principles.

Flavour diagonal mesons are difficult to study on the lattice for a variety of
reasons: Firstly, disconnected quark line diagrams contribute substantially
to the correlation functions which, as a result, are noisy and therefore
computationally demanding. Secondly, the $\eta$ and $\eta^{\prime}$ mesons are
no flavour eigenstates and require a set of multiple interpolators to
create them efficiently. In the literature this is often referred to as the
``mixing'' of flavour octet and singlet ``states'' (although there is no
mixing in QCD of mass eigenstates). A careful analysis and sophisticated methods
are required to isolate the respective ground state contributions
in the presence of higher excited states from the resulting matrix of
correlation functions. Only a precise knowledge of the resulting
linear combinations of interpolators enables us to compute
matrix elements, where these states are destroyed by axial and
pseudoscalar local quark bilinears or by local gluonic operators
and to determine the decay constants and anomaly matrix elements.

Although there have been previous lattice determinations of the masses~\cite{Kuramashi:1994aj,Venkataraman:1997xi,Bardeen:2000cz,Struckmann:2000bt,McNeile:2000hf,Bali:2001gk,Lesk:2002gd,Hashimoto_2008,Jansen:2008wv,Christ:2010dd,Dudek:2011tt,Gregory:2011sg,Bali:2014pva,Fukaya:2015ara,Sun:2017ipk,Dimopoulos:2018xkm,Kordov:2021eqx} and pseudoscalar matrix elements~\cite{Ottnad:2012fv,Michael:2013gka,Bali:2014pva,Ottnad:2017bjt} that can be related to the decay constants within certain model assumptions, a thorough physical point determination of the decay constants and anomalous matrix elements has only recently been published by the present authors~\cite{Bali:2021qem}. Here we summarize these results that have been obtained, carefully extrapolating to the continuum limit and employing next-to-leading-order (NLO) large-$N_{c}$ ChPT in the continuum. In addition we discuss the role of higher lying states whose structure in terms of singlet and octet contributions is not well known phenomenologically.

\section{Lattice Setup}
\begin{figure}
  \includegraphics[width=\linewidth]{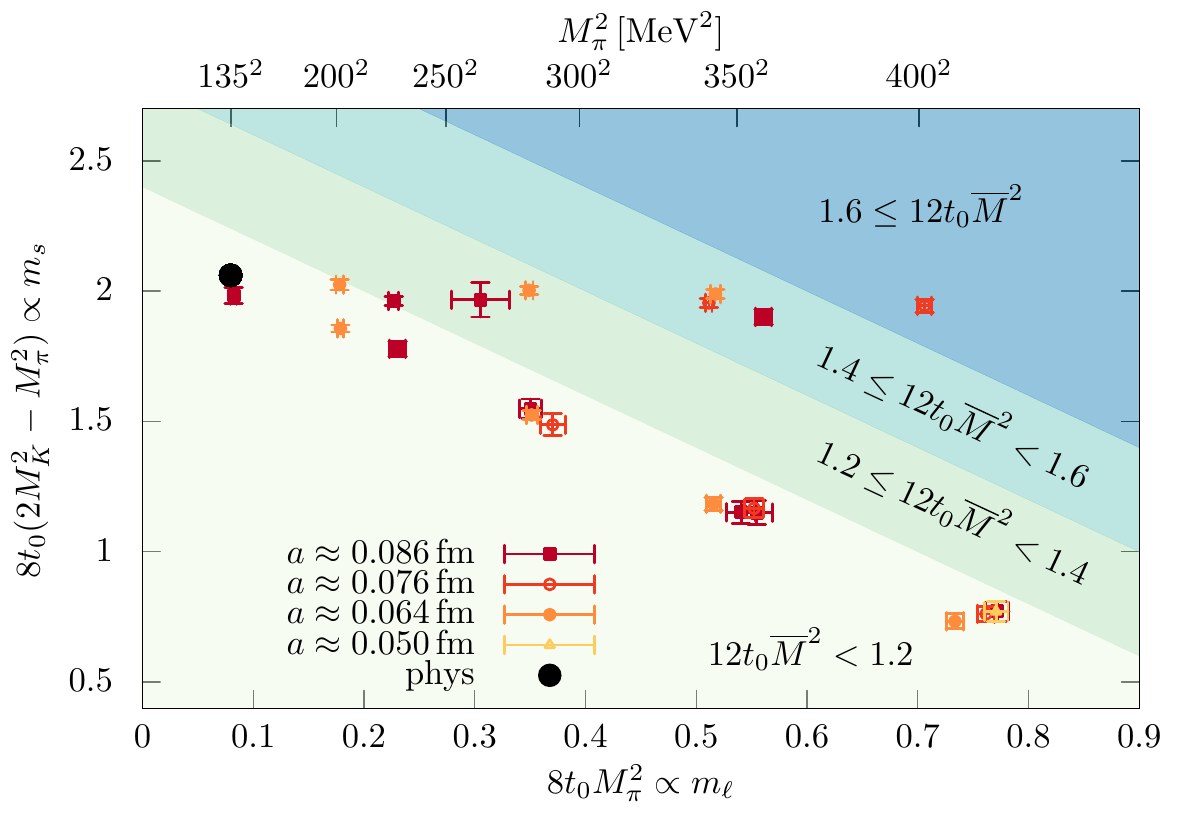}
  \caption{The positions of the analysed CLS ensembles in the quark mass plane. Both trajectories intersect close to the physical point (black circle) and one ensemble (D150) is very close to that point. Shading represent cuts on the average pseudoscalar meson mass $12 t_{0}\overline{M}^{2} = 4 t_{0} (2 M_{K}^{2} + M_{\pi}^{2})$ that we impose during our chiral extrapolation procedure. Symbols encode the four lattice spacings.
    \label{fig:quarkmasses}}
\end{figure}
We employ twenty-one CLS gauge ensembles
generated with $N_{f}=2+1$ non-perturbatively $\mathcal{O}(a)$
improved Wilson fermions. For details, see~\cite{Bruno:2014jqa}.
Two mass trajectories were realized that both lead to
the physical point. Along one trajectory the average
quark mass is held fixed~\cite{Bruno:2014jqa},
along the other trajectory the strange
quark mass is kept approximately constant~\cite{fixeds}.
All ensembles have large volumes with $L_{s} > 2.2\,{\rm fm}$ and $L_{s}M_{\pi} > 4$ on most of them. To have full control over the continuum limit extrapolation, we incorporate four lattice spacings, $0.050\,{\rm fm} \le a \le 0.086\,{\rm fm}$.

On every configuration, we compute a matrix of correlation functions,
\begin{align}
  \label{eq:matcor}
  C_{ij}(\vec{p},t)=\frac{1}{N_{t_{\mathrm{in}}}}\sum_{t_{\mathrm{in}}}
  \left\langle\Omega\left|\mathcal{B}_i(\vec{p},t+t_{\mathrm{in}})
  \mathcal{B}^{\dagger}_j(-\vec{p},t_{\mathrm{in}})\right|\Omega\right\rangle,
\end{align}
where $|\Omega\rangle$ is the vacuum and $N_{t_{\mathrm{in}}}$ denotes
the number of source time slices
that we average over. We mostly use open boundary conditions and in these cases we sum only over the bulk of the lattice where boundary effects are negligible.

The interpolators $\mathcal{B}_{i}$ are octet and singlet combinations of
spatially smeared pseudoscalar quark-antiquark interpolating operators,
\begin{align}
  \mathcal{P}^8&=\frac{1}{\sqrt{6}}\left(\mathcal{P}^u+\mathcal{P}^d-2\mathcal{P}^s\right)=\frac{1}{\sqrt{3}}\mathcal{P}^{\ell}
  -\sqrt{\frac{2}{3}}\mathcal{P}^s,\label{eq:p8}\\
  \mathcal{P}^0&=\frac{1}{\sqrt{3}}\left(\mathcal{P}^u+\mathcal{P}^d+\mathcal{P}^s\right)=\sqrt{\frac{2}{3}}\mathcal{P}^{\ell}+\frac{1}{\sqrt{3}}\mathcal{P}^s\label{eq:p0}
\end{align}
with $\mathcal{P}^{q} = \overline{q}\gamma_{5}q$, $q\in\{u,d,s\}$ and  $\mathcal{P}^{\ell} = \frac{1}{\sqrt{2}}\left(\overline{u}\gamma_{5}u + \overline{d}\gamma_{5}d\right)$.
To extract matrix elements, we also use corresponding (partially $\mathcal{O}(a)$-improved) local
axialvector $A_\mu^a=\bar{\psi}t^a\gamma_\mu\gamma_5\psi + a c_{A} \partial_{\mu}P^{a}$
($\bar{\psi} =(\bar{u},\bar{d},\bar{s})$), and pseudoscalar currents $P^a= \bar{\psi}t^a\gamma_5\psi$ at the sink.
Note that this normalization
differs from that of eqs.~\eqref{eq:p8}--\eqref{eq:p0} by a factor $1/\sqrt{2}$.
The disconnected loops entering the correlators are estimated using 96
stochastic sources. To keep near-neighbour noise under control we subtract
the (for our action and operator) maximum possible
number of terms of the hopping parameter expansion~\cite{Bali:2009hu} from the
propagator and place the stochastic sources at every fourth timeslice
 (dilution, partitioning~\cite{Bernardson:1993he}) at a time. This corresponds to about 300 to 600 thousand solutions of the Wilson-Dirac equation per ensemble. For the extended interpolators, we employ two different levels of smearing. The matrix of correlators is usually built from $\mathcal{B}_{i}\in\{\mathcal{P}^{\ell}, \mathcal{P}^{s}, \mathcal{P}^{8}\}$ at two different smearing levels, except at the symmetric $m_s=m_{\ell}$ point, where the Wick contraction between singlet and octet currents vanishes and yields a block-diagonal matrix. This enables us to split the problem into two parts and to analyse pure octet and singlet correlation matrices separately.

Instead of solving the generalized eigenvalue problem~\cite{Michael:1985ne,Luscher:1990ck}, we directly fit to the Euclidean time derivative to improve the signal~\cite{Feng:2009ij,Umeda:2007hy},
\begin{equation}
  \label{eq:dtrick}
  \partial_t C(t) \sim Z\left( \partial_t D(t) \right)Z^{\intercal},
\end{equation}
where $\partial_t C(t) = \left( C(t+a) - C(t-a) \right)/(2a)$ is the
symmetric discretized derivative and
\begin{equation}
  \partial_t D(t) = -\mathrm{diag}\left[ E_n \exp\left(-E_n t\right) \right]
\end{equation}
for open boundaries. For periodic boundaries the back-propagating part is taken into account. All time-dependence is contained in the diagonal matrix of eigenvalues, $D$, while the amplitudes are encoded in the matrix $Z_{in} = \langle \mathcal{B}_{i}| n \rangle / \sqrt{2 E_{n} V_{3}}$, where $|n\rangle$ is the eigenstate corresponding to the $n$-th energy level $E_{n}$ and $V_{3}$ is the spatial volume.

\begin{figure}[htp]
  \centering
  \resizebox{0.98\linewidth}{!}{\input{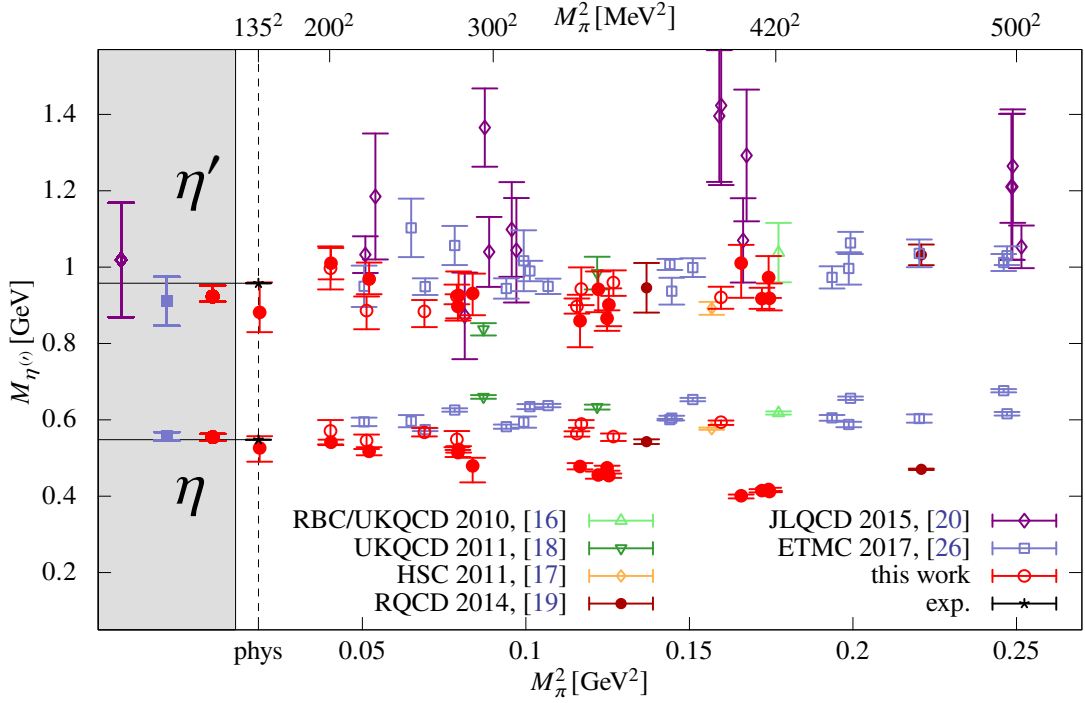}}
  \caption[Masses]{Recent $N_f=2+1(+1)$ lattice results for the masses of the $\eta$
    and $\eta^\prime$ mesons. Most points have been simulated at approximately
    physical strange quark masses (open symbols), whereas in this work we also include an additional
    trajectory along which the average of the quark masses is kept constant (filled
    symbols). The three sets of points in the shaded regions left of the physical point (dashed line)
    correspond to the continuum and chirally extrapolated results of JLQCD~\cite{Fukaya:2015ara} (who do not give an estimate of $M_{\eta}$), ETMC~\cite{Ottnad:2017bjt} and this work. \label{fig:masssummary}
  }
  \label{fig:masses}
\end{figure}

To improve the stability of the multi-exponential fit and to increase the sensitivity to the $\eta^{\prime}$ state, we include a generalized version of the effective mass into our fully correlated fit,
\begin{align}
  \partial_t\log C(t) = & (\partial_t C(t))C^{-1}(t)
  \approx  \left( {Z} \partial_t {D}(t) {Z}^{\intercal} \right)\left( {Z} {D}(t) {Z}^{\intercal}\right)^{-1}
  = - Z \, \mathrm{diag}_{n=0}^{N-1}(E_n)\, Z^{-1},
  \label{eq:generalizedeffectivemass}
\end{align}
which is
constant in time~(up to excited states corrections and statistical noise).
We further improve the signal by including data at the first non-zero momentum into our combined fit, employing the continuum dispersion relation,
$E_{n}(\vec{p}) = \sqrt{M_{n}^{2} + a^{2} \vec{p}^{2}}$. We show the results on the individual ensembles together with a comparison to other lattice determinations in fig.~\ref{fig:masses}

Having extracted energies and eigenstates of the $\eta$ and $\eta^{\prime}$ mesons,
we are in a position to compute decay constants,
\begin{align}
  \label{eq:defDecConst}
  Z_{A}^{a}(a,\mu) \left\langle \Omega\left|A_{\mu}^a\right| \mathcal{M}(p)\right\rangle=iF^a_{\mathcal{M}}p_{\mu},
\end{align}
where $Z_{A}^{a}(\mu)$ is a renormalization factor. While $Z_A=Z_A^8$
is scale-independent, the singlet renormalization factor
$Z_A^s(\mu)=Z_A^0(\mu)$ depends on the scale. The normalization
$A_{\mu}^a=\bar{\psi}t^a\gamma_\mu\gamma_5\psi$ that we adopt corresponds
to $F_{\pi^0}^3\approx 92\,$MeV. Note that there are two decay constants each
for $\mathcal{M}\in\{\eta,\eta'\}$: singlet ($a = 0$) and octet
($a = 8$). These are often parametrized in terms of two decay constant
parameters $F^0$ and $F^8$ and two angles $\theta_{8}$ and $\theta_{0}$,
\begin{align}
  \label{eq:octsingletanglerep}
\begin{pmatrix}
  F_\eta^8 & F_\eta^0\\   F_{\eta^\prime}^8 & F_{\eta^\prime}^0
\end{pmatrix}
=
\begin{pmatrix}
  F^8 \cos \theta_8 & - F^0 \sin \theta_0 \\   F^8 \sin \theta_8 & F^0\cos \theta_0
\end{pmatrix}.
\end{align}
\section{Physical point results for masses and decay constants}
We perform a combined continuum limit and chiral extrapolation of both masses and the four decay constants, employing large-$N_{c}$ ChPT to NLO~\cite{Gasser:1984gg,Bickert:2016fgy}. This allows us to describe the quark mass dependence of these six observables in the continuum theory by a common set of six low-energy constants: the pion decay constant in the chiral limit, $F$, the large-$N_{c}$ versions of the familiar SU(3) LECs $L_{5}$ and $L_{8}$, as well as the LECs $M_0^2$, $\Lambda_1$ and $\Lambda_2$. The singlet mass $M_{0}(\mu)$ in the chiral limit is related to the topological susceptibility and depends on the QCD scale $\mu$. The remaining two OZI suppressed parameters $\Lambda_{1}(\mu)$ and $\Lambda_{2}(\mu)$ depend on the
QCD renormalization scale too. This is due to the anomalous dimension of the
singlet axialvector current. Since this vanishes at leading order, in the limit $\mu\rightarrow\infty$, the singlet renormalization factor $Z_{A}^{s}(\infty)$ remains finite and we perform our fits in that limit. If needed, the results can then be evolved back to lower energy scales. The difference between the singlet and the non-singlet renormalization factors for our lattice action is known perturbatively to two loops~\cite{Constantinou:2016ieh} and we use the non-perturbatively determined $Z_{A}$~\cite{Bulava:2016ktf} for the non-singlet current. Note that chiral logs only appear at NNLO in the large-$N_c$ ChPT power counting. Therefore, all NLO LECs are independent of the ChPT renormalization scale (but some depend on the QCD scale $\mu$).

In order to account for the lattice spacing dependence, we include parameterizations of three unknown linear improvement coefficients of the singlet and one of the octet current into our fit, while setting the remaining coefficients to their literature values~\cite{improve2}. In this way, full $\mathcal{O}(a)$ improvement is achieved with four additional fit parameters.
We then add systematically $\mathcal{O}(a^{2}\Lambda^{2})$, $\mathcal{O}(a^{2}\Lambda(2m_{\ell} + m_{s}))$ and $\mathcal{O}(a^{2}\Lambda(m_{s} - m_{\ell}))$ terms to all observables and subsequently remove those that we cannot determine.

\begin{figure}[htp]
  \centering
  \includegraphics[width=0.9\linewidth]{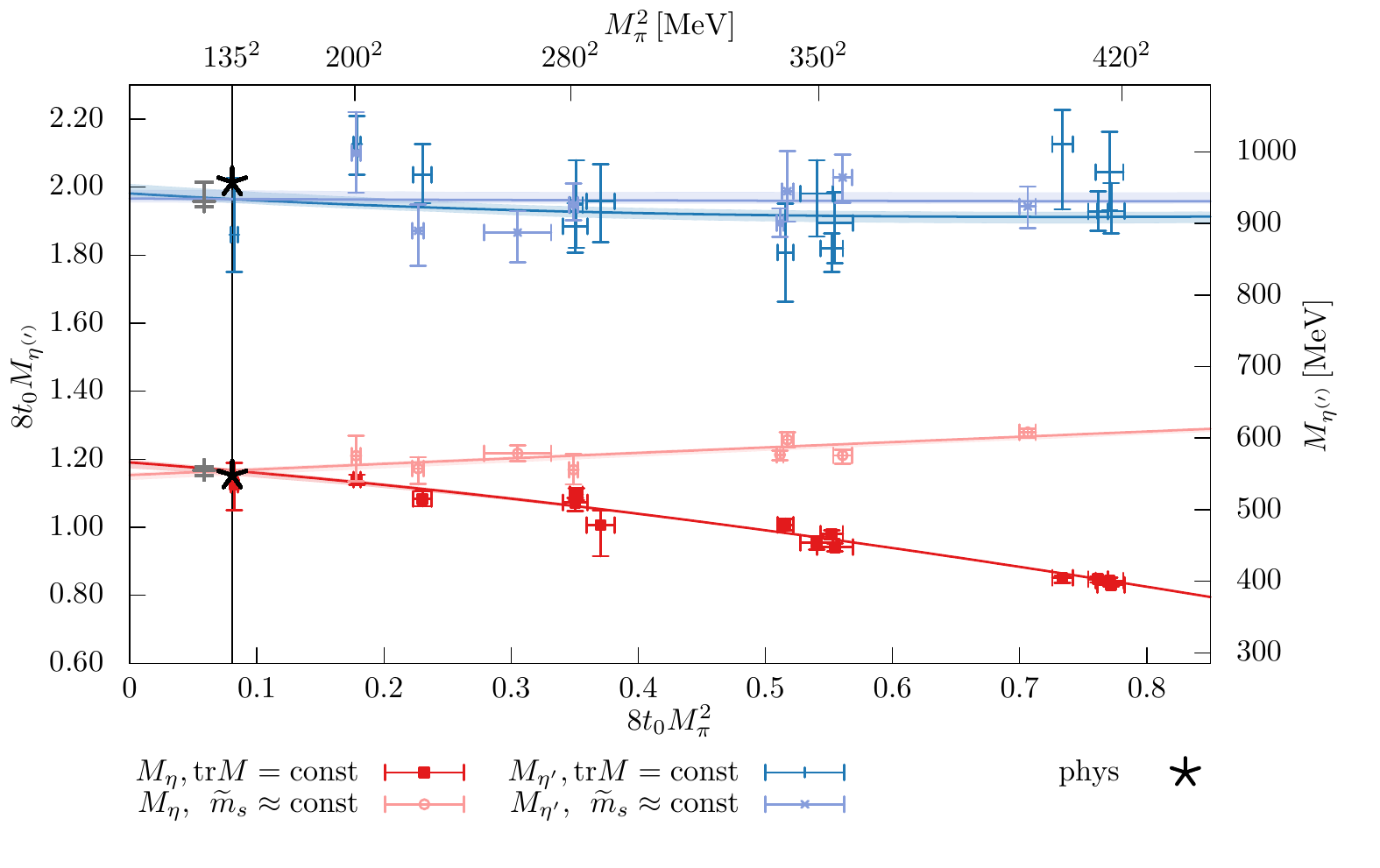}
  \includegraphics[width=0.9\linewidth]{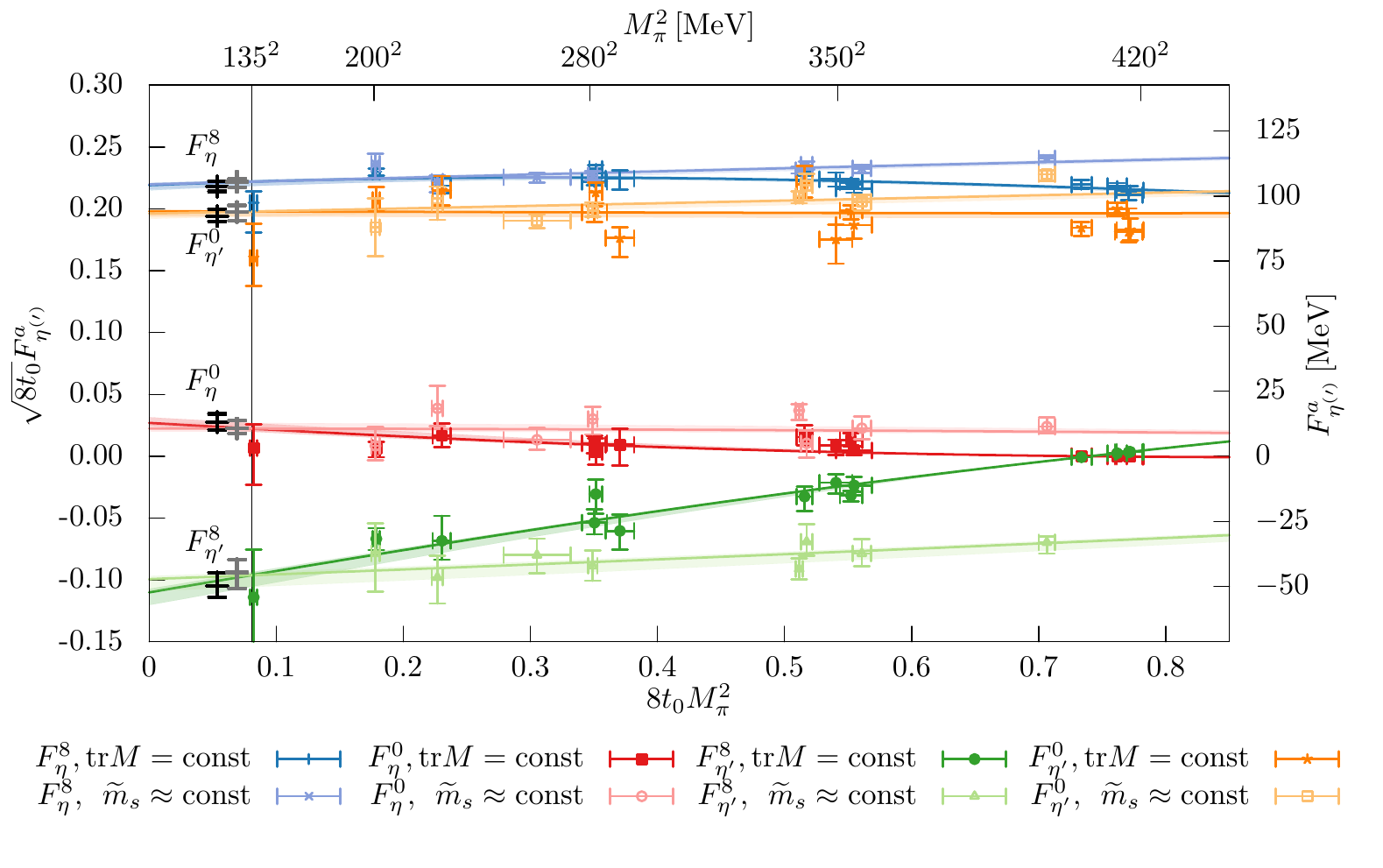}
  \caption[Chiral extrapolation]{Chiral extrapolation of the masses (top) and decay constants (bottom) as a function of $M_\pi^2$. The two analyzed trajectories meet at the physical point.}
  \label{fig:chiralextrap}
\end{figure}

In this way, we carried out 17 different fits that all gave acceptable $\chi^{2}$ values. The systematic error associated with the continuum limit is computed from the central 68.5\,\% range of the scatter of these results. On every ensemble we take into account the correlations between the non-singlet quark masses (the arguments of the fit functions) and the observables using Orear's method~\cite{Orear:1981qt}. The best fit gives $\chi^{2} / N_{\rm df} \approx 1.47$ and we quote its result as central values. We plot all our results at non-physical quark masses and their extrapolations to the physical point in fig.~\ref{fig:chiralextrap}.

To estimate the systematic uncertainties from the chiral extrapolation we follow a similar approach and remove ensembles with large average squared non-singlet pseudoscalar masses $12 t_{0}\overline{M}^{2} = 4 t_{0}(2 M_{K}^{2} + M_{\pi}^{2})$, see fig.~\ref{fig:quarkmasses}. We also repeat fits for different initial energy scales $\mu_{0}\in\{a^{-1}/2, a^{-1}, 2a^{-1}\}$ from which we start the evolution of the scale-dependent singlet renormalization factor to $\mu = \infty$.

\begin{table*}
  \centering
  \begin{tabularx}{\textwidth}{l|l|l}
    \toprule
    & no priors & exp. masses as priors \\
    \toprule
    $L_5$ & $\phantom{-}1.58\left(\substack{17\\7}\right)_\mathrm{stat} \left(\substack{0\\22}\right)_a\left(\substack{20\\9}\right)_\chi\cdot 10^{-3}$ & $\phantom{-}1.66\left(\substack{12\\9}\right)_\mathrm{stat} \left(\substack{0\\26}\right)_a\left(\substack{13\\8}\right)_\chi\cdot 10^{-3}$\\\midrule
  $L_8$ & $\phantom{-}0.96\left(\substack{15\\6}\right)_\mathrm{stat} \left(\substack{0\\14}\right)_a\left(\substack{16\\7}\right)_\chi\cdot 10^{-3}$ & $\phantom{-}1.08\left(\substack{11\\6}\right)_\mathrm{stat} \left(\substack{0\\12}\right)_a\left(\substack{3\\10}\right)_\chi\cdot 10^{-3}$\\\midrule
  $M_0(\mu=\infty)$ & $\phantom{-}1.67\left(\substack{2\\6}\right)_\mathrm{stat} \left(\substack{1\\2}\right)_a\left(\substack{0\\1}\right)_\chi\left(\substack{3\\2}\right)_{\mathrm{renorm}} \left(8t_0^{\chi}\right)^{-1/2}$ & $\phantom{-}1.62\left(\substack{2\\4}\right)_\mathrm{stat} \left(\substack{3\\1}\right)_a\left(\substack{3\\0}\right)_\chi\left(\substack{2\\1}\right)_{\mathrm{renorm}}\left(8t_0^{\chi}\right)^{-1/2}$\\
  & $= 785\left(\substack{9\\28}\right)_\mathrm{stat} \left(\substack{15\\14}\right)_{\mathrm{syst}}(12)_{t_0}\,\mathrm{MeV}$& $= 761\left(\substack{13\\21}\right)_\mathrm{stat} \left(\substack{18\\11}\right)_{\mathrm{syst}}(11)_{t_0}\,\mathrm{MeV}$\\\midrule
  $F$ & $\phantom{-}0.1890\left(\substack{27\\37}\right)_\mathrm{stat} \left(\substack{36\\0}\right)_a\left(\substack{21\\38}\right)_\chi\left(8t_0^{\chi}\right)^{-1/2}$ & $\phantom{-}0.1866 \left(\substack{26\\29}\right)_\mathrm{stat} \left(\substack{54\\0}\right)_a\left(\substack{19\\16}\right)_\chi\left(8t_0^{\chi}\right)^{-1/2}$\\
  & $= 88.83\left(\substack{1.27\\1.74}\right)_\mathrm{stat} \left(\substack{1.96\\1.79}\right)_{\mathrm{syst}}(1.32)_{t_0}\,\mathrm{MeV}$ & $= 87.71\left(\substack{1.44\\1.57}\right)_\mathrm{stat} \left(\substack{2.69\\81}\right)_{\mathrm{syst}}(1.31)_{t_0}\,\mathrm{MeV}$\\\midrule
  $\Lambda_1(\mu=\infty)$ & $-0.22\left(\substack{1\\5}\right)_\mathrm{stat} \left(\substack{0\\3}\right)_a\left(\substack{0\\3}\right)_\chi\left(\substack{6\\3}\right)_{\mathrm{renorm}}$ & $-0.25\left(\substack{1\\4}\right)_\mathrm{stat} \left(\substack{3\\1}\right)_a\left(\substack{1\\1}\right)_\chi\left(\substack{5\\2}\right)_{\mathrm{renorm}}$\\
  $\Lambda_2(\mu=\infty)$ & $-0.1\left(\substack{8\\4}\right)_\mathrm{stat} \left(\substack{0\\10}\right)_a\left(\substack{14\\8}\right)_\chi\left(\substack{5\\3}\right)_{\mathrm{renorm}}$ & $\phantom{-}0.11\left(\substack{5\\5}\right)_\mathrm{stat} \left(\substack{0\\9}\right)_a\left(\substack{6\\5}\right)_\chi\left(\substack{3\\2}\right)_{\mathrm{renorm}}$\\
    \bottomrule
  \end{tabularx}
  \caption{Large-$N_{c}$ LECs parametrizing the continuum limit of our results on the masses and decay constants. The left column is obtained from a unconstrained but fully correlated fit to all 21 ensembles. The right column is a fit to the same data but adding the physical masses~\eqref{eq:physetaetaprimemasses} as a prior and gives very similar results.
    The conversion to physical unit was done using $(8 t_{0}^{\chi})^{-1/2} =470(7)\,\mathrm{MeV}$ in the chiral limit~\cite{Bruno:2014jqa,spectrum,Bali:2021qem}.
    \label{tab:lecs}}
\end{table*}

Our central fit is parameterized in the continuum by the LECs listed in the left column of tab.~\ref{tab:lecs} and yields for the masses at the physical point,
\begin{align}
  M_\eta &=
  1.168\left(\substack{8\\14}\right)_\mathrm{stat} \left(\substack{1\\0}\right)_a\left(\substack{5\\6}\right)_\chi
  (8t_0^{\rm ph})^{-1/2}
  =
  554.7\left(\substack{4.0\\6.6}\right)_\mathrm{stat} \left(\substack{2.4\\2.7}\right)_{\mathrm{syst}}(7.0)_{t_0} \,\mathrm{MeV}
  \quad\text{and}\\
  M_{\eta^\prime} &=
  1.958\left(\substack{27\\13}\right)_\mathrm{stat} \left(\substack{0\\6}\right)_a\left(\substack{48\\3}\right)_\chi
(8t_0^{\rm ph})^{-1/2}
    = 929.9\left(\substack{12.9\\6.0}\right)_\mathrm{stat} \left(\substack{22.9\\3.3}\right)_{\mathrm{syst}}(11.7)_{t_0}
  \,\mathrm{MeV},
\label{eq:massfitresults}
\end{align}
using the Wilson flow scale at the physical point $(8t_{0}^{\rm ph})^{-1/2} = 475(6)\,\mathrm{MeV}$~\cite{Bruno:2016plf} to convert to physical units.
We find reasonably good agreement when
comparing these results of $N_{f} = 2+1$ QCD with the known experimental masses,
\begin{equation}
  \text{PDG\,\cite{PDG}}:\qquad
  M^{\mathrm{ph}}_\eta = 547.862(17)\,\mathrm{MeV}
  \quad\text{and}\quad
  M^{\mathrm{ph}}_{\eta^\prime} = 957.78(6)\,\mathrm{MeV}\label{eq:physetaetaprimemasses}.
\end{equation}
The masses are 0.7 standard errors above and one standard error below
the experimental values for the $\eta$ and $\eta^{\prime}$,
respectively.
We also carry out a second fit incorporating the knowledge of the physical masses by adding them as priors to the $\chi^{2}$ function and obtain similar results for the LECs listed in the right column of tab.~\ref{tab:lecs}. Using this second set of LECs, we obtain for the decay constants
in the angle representation, eq.~\eqref{eq:octsingletanglerep},
\begin{align}
      F^8 &= 0.2421\left(\substack{22\\26}\right)_\mathrm{stat} \left(\substack{8\\50}\right)_a\left(\substack{32\\12}\right)_\chi\,(8 t_0^{\rm ph})^{-1/2}
= 115.0\left(\substack{1.1\\1.2}\right)_\mathrm{stat} \left(\substack{1.6\\2.4}\right)_{\mathrm{syst}}(1.5)_{t_0}\,\mathrm{MeV},\\
      \theta_8 &= -0.450\left(\substack{21\\36}\right)_\mathrm{stat} \left(\substack{24\\0}\right)_a\left(\substack{29\\0}\right)_\chi\left(\substack{1\\5}\right)_{\mathrm{renorm}} = -25.8\left(\substack{1.2\\2.1}\right)_{\mathrm{stat}}\left(\substack{2.2\\0.3}\right)_{\mathrm{syst}} ^\circ,\\
  F^0(\mu = 2\,\mathrm{GeV}) &= 0.2108\left(\substack{14\\40}\right)_\mathrm{stat} \left(\substack{12\\25}\right)_a\left(\substack{30\\8}\right)_\chi\left(\substack{50\\26}\right)_{\mathrm{renorm}}\,(8 t_0^{\rm ph})^{-1/2}\nonumber\\
&= 100.1\left(\substack{7\\1.9}\right)_\mathrm{stat} \left(\substack{2.0\\2.7}\right)_{\mathrm{syst}}(1.3)_{t_0}\,\mathrm{MeV},\\
      \theta_0 &= -0.141\left(\substack{18\\20}\right)_\mathrm{stat} \left(\substack{0\\27}\right)_a\left(\substack{27\\0}\right)_\chi = -8.1\left(\substack{1.0\\1.1}\right)_{\mathrm{stat}}\left(\substack{1.5\\1.5}\right)_{\mathrm{syst}}^\circ.
\end{align}
In comparison to the frequently used flavour representation, the octet/singlet representation has the advantage that only the singlet decay constants depend on the renormalization scale. We refer to tabs.~24 and~25 of~\cite{Bali:2021qem} for results at several scales and in different parameterizations.

\begin{figure}
  \centerline{\includegraphics[width=0.6\linewidth]{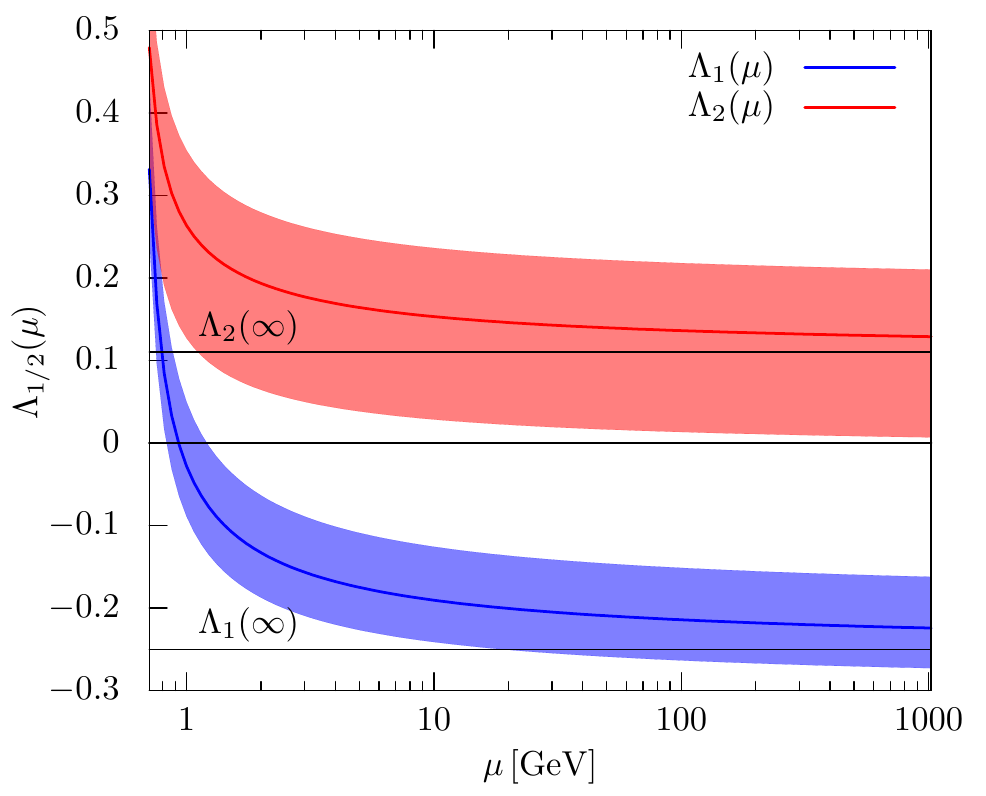}}
  \caption{Scale dependence of the large-$N_{c}$ ChPT LECs $\Lambda_{1}$ and $\Lambda_{2}$.
    \label{fig:lambdasAtScale}}
\end{figure}

The QCD scale-dependence cannot be determined from phenomenological fits to
experimental low-energy data. Therefore, only scale-independent combinations
of the large-$N_{c}$ ChPT LECs $M_{0}^{2}, \Lambda_{1}$ and
$\Lambda_{2}$~\cite{Leutwyler:1997yr} are accessible, unless additional
assumptions are made. The most prominent framework is the Feldmann-Kroll-Stech
(FKS) model~\cite{Feldmann:1998vh,Feldmann:1998sh,Feldmann:1999uf} which uses
the fact that in the flavour basis the difference between the light and
strange mixing angles is OZI suppressed and proportional to $\Lambda_{1}$.
This reduces the number of independent parameters for the four decay
constants to three, setting $\phi_{\ell} = \phi_{s}$ and assuming
$\Lambda_{1} = 0$. Our analysis shows that the latter approximation
indeed holds at low energy scales ($\mu \approx 1\,\mathrm{GeV}$),
see fig.~\ref{fig:lambdasAtScale}, and that the FKS model works to
a reasonable accuracy. Therefore, our results are in good agreement
with~\cite{Feldmann:1999uf} and many other phenomenological determinations.

\section{Pseudoscalar states above the $\eta^{\prime}$}
\begin{figure}[thp]
  \centering
  \includegraphics[width=0.8\linewidth]{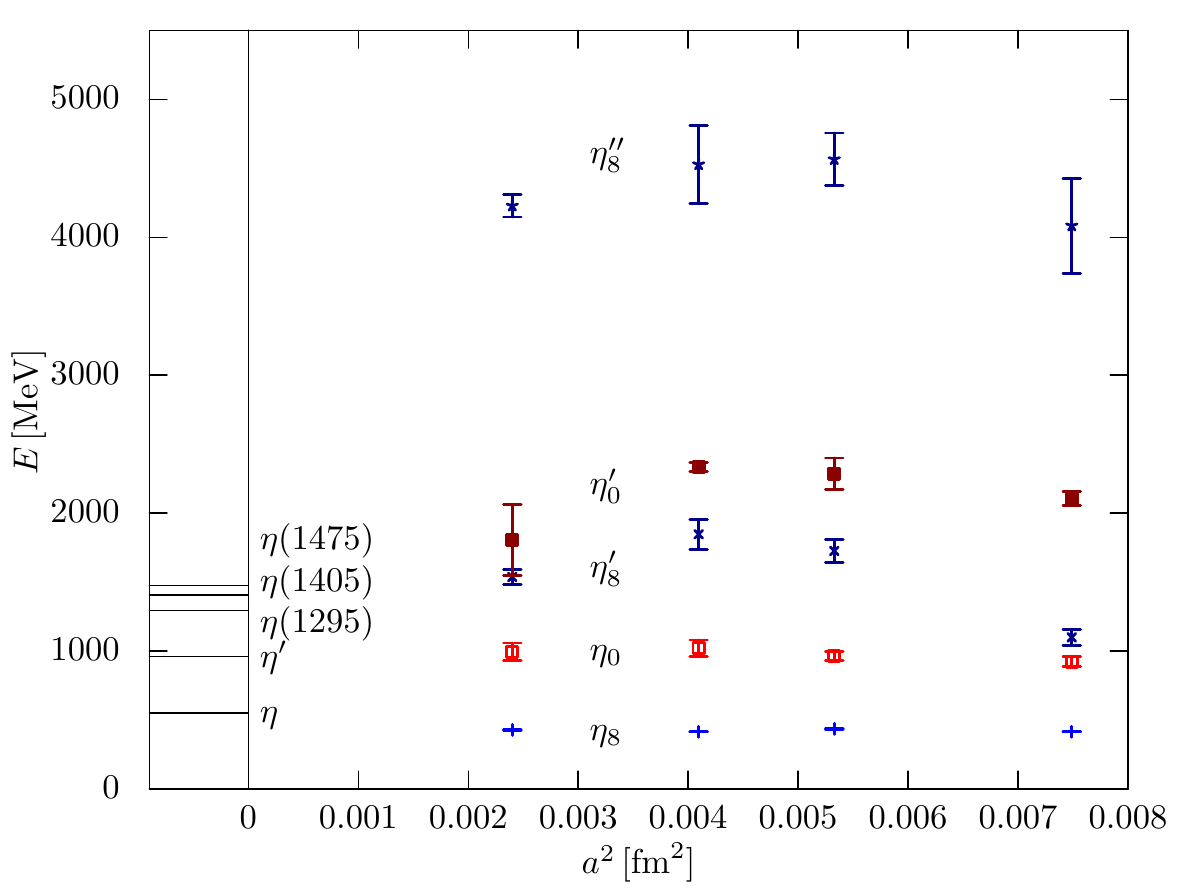}
  \caption[Excited states]{The octet (blue) and singlet (red)
    energy levels at the $m_s=m_{\ell}$ point versus
    the squared lattice spacing.
    For orientation also the experimental resonances (at physical quark masses)
    are shown.\label{fig:excitedstates}}
  \label{fig:excstates}
\end{figure}
We usually analyse a $3\times 3$ matrix of correlation functions
to extract the lowest two energy levels, the $\eta$ and the $\eta^{\prime}$.
The higher excitations are not resolved but effectively described by a
third energy level.
At the $N_f=3$ symmetric point the matrix of correlation functions
is block diagonal regarding the singlet and octet sectors.
Therefore, at this point, we can resolve higher excitations.
In experiment several resonances with the same quantum numbers
are relatively close to the $\eta^{\prime}$, namely the
$\eta(1295)$, $\eta(1405)$ and the $\eta(1475)$. The first state
decays mostly into $\eta\pi\pi$. This may indicate that the octet component
dominates. In contrast, $\eta(1405)$ and $\eta(1475)$ decay into
$K\bar{K}\pi$, $a_0\pi$ etc. The $\eta(1405)$ which, unlike
the $\eta(1475)$, has not been observed to decay into $\gamma\gamma$
has even been postulated to be a glueball~\cite{Cheng:2008ss}, however,
unlike the $\eta(1475)$ and like the $\eta(1295)$,
it also decays into $\eta\pi\pi$.

On our $m_s=m_{\ell}$ ensembles we use a $2\times 2$ matrix of smeared singlet interpolators for the $\eta^{\prime} = \eta_{0}$ and in case of the octet we
include the local current in addition, resulting in a $3 \times 3$ matrix.
This enables us to determine the first excited singlet and the first two
excited octet states, see fig.~\ref{fig:excstates}.
At $a=0$ we have included the experimental spectrum
for orientation.
One should keep in mind, however, that this refers to the physical
point and not to $m_s=m_{\ell}$.
We observe large cutoff effects that have a similar shape for all the
excited states. In contrast, for the ground states we are unable to
resolve any lattice spacing dependence.
The third octet state is around 4\,GeV, indicating that the first excitation
becomes the $\eta(1295)$ at the physical point and that
the $\eta(1405)$ and $\eta(1475)$ may have large singlet components,
although we are unable to resolve more than one singlet state within
this energy region within our statistical accuracy.

\section{Gluonic matrix elements}
\begin{figure}
  \includegraphics[width=0.49\linewidth]{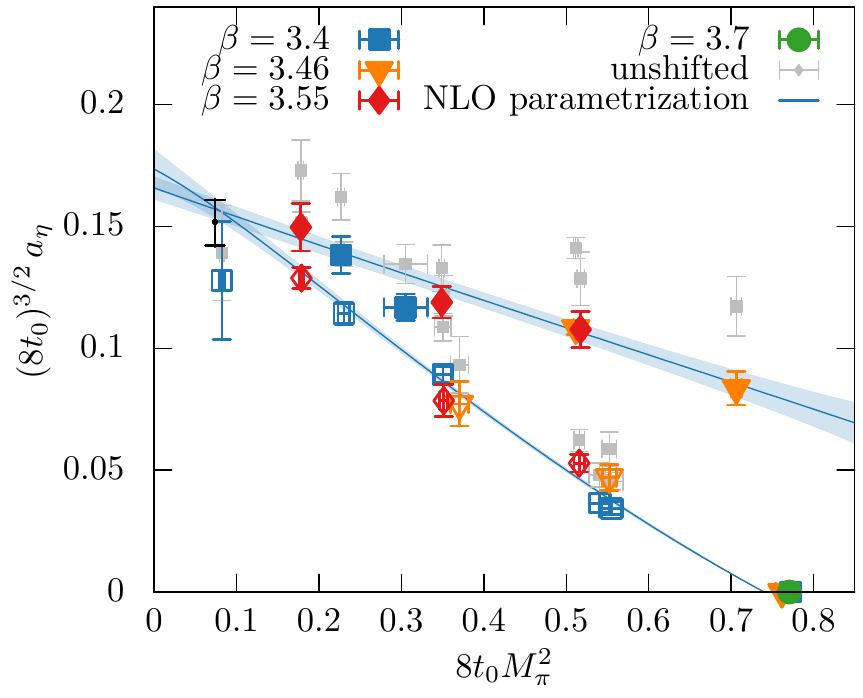}
  \includegraphics[width=0.49\linewidth]{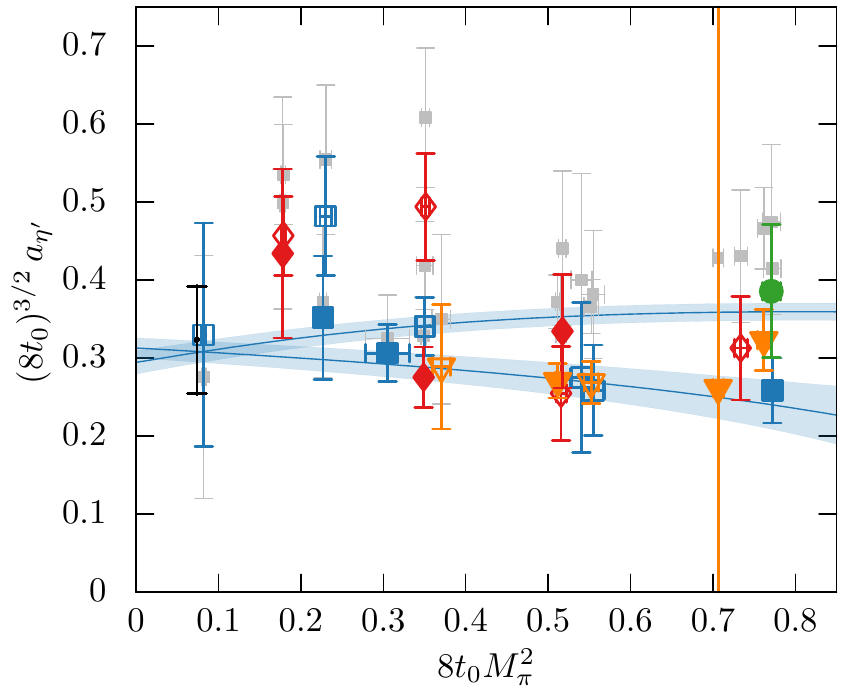}
  \caption{\label{fig:gluonme}The anomalous gluonic matrix element
    $a_\eta$ (left) and $a_{\eta^\prime}$ (right)
    determined via the singlet AWI from fermionic matrix elements,
    eq.~\eqref{eq:gluonmefermionic}. The coloured points have been
    adjusted for lattice spacing effects, while the grey points
    indicate the unshifted data. The two curves correspond to the NLO
    large-$N_c$ ChPT parametrization derived in~\cite{Bali:2021qem}
    for trajectories with a constant average quark mass and a constant
    strange quark mass.  The black error bars indicate the final results at the physical
  point including statistical and systematic errors.}
\end{figure}

After constructing interpolators that create the $\eta$ and $\eta'$ mesons,
based on the overlap factors $Z_{in}$ of eq.~\eqref{eq:dtrick},
we may also compute local matrix elements other than the
decay constants.
The singlet axial Ward identity connects axial and pseudoscalar quark
bilinear currents with the topological charge density
$\omega(x)=-\tfrac{1}{32\pi^2}F_{\mu\nu}^a(x)\widetilde{F}_{\mu\nu}^a(x)$,
\begin{equation}
  \partial_{\mu}\widehat{A}^{a}_{\mu}=\widehat{\left(\overline{\psi}\gamma_5\{M,t^a\}\psi\right)}
  +\sqrt{6}\delta^{a0}\widehat{\omega},
  \label{eq:awi}
\end{equation}
where $M=\mathrm{diag}(m_\ell,m_\ell,m_s)$
is the quark mass matrix, $a\in\{0,\ldots,8\}$ labels the generators of U(3), $t^{a} = \lambda^{a}/2$ and $t^0 = \mathds{1}/\sqrt{6}$. Hats denote renormalized quantities. For the determination of the topological charge
density we evolve the gauge fields to a gradient flow time
$\sqrt{8t} = \sqrt{8 t_{0}^{*}} \approx 0.413\,\mathrm{fm}$.
Under renormalization $\omega$ mixes with $\partial_{\mu} A_{\mu}^{0}$,
\begin{equation}
  \label{eq:renormomega}
  \hat{\omega} = Z_{\omega} \omega + Z_{\omega A}\partial_{\mu} A_{\mu}^{0},
\end{equation}
and the renormalization factors are unknown. We therefore start by constructing the
anomalous matrix elements $a_{\eta^{(\prime)}} = 2 \langle \Omega | \hat{\omega} | \eta^{(\prime)}\rangle$ by
combining axialvector and pseudoscalar matrix elements,
\begin{equation}
  a_{\mathcal{M}}(\mu)=
    \sqrt{\frac{2}{3}}Z_A^{s}(\mu)\partial_{\mu}\left\langle
    \Omega\left|A_{\mu}^0\right|\mathcal{M}\right\rangle
    +\frac{2\sqrt{2}}{\sqrt{3}}Z_A\left[\frac{\sqrt{2}}{3}\delta\widetilde{m}
      \left\langle
    \Omega\left|P^8\right|\mathcal{M}\right\rangle
    -r_P\overline{\widetilde{m}}\left\langle
    \Omega\left|P^0\right|\mathcal{M}\right\rangle
    \right].
    \label{eq:gluonmefermionic}
\end{equation}
The PCAC masses $\overline{\widetilde{m}} = (2 \widetilde{m}_{\ell} + \widetilde{m}_{s})/3$ and $\delta\widetilde{m} = \widetilde{m}_{s}-\widetilde{m}_{\ell}$ are determined from the non-singlet PCAC relations (e.g., $a=1$ and $a=4$) and
we set $r_{P} = Z_{P}^{s}/Z_{P} = 1$ since the difference $Z_{P}^{s} - Z_{P}=\mathcal{O}(g^{6})$ for Wilson quarks.
We then attempt a combined fit to the $a_{\eta}$ and $a_{\eta^{\prime}}$ data, again parameterizing all four additional unknown pseudoscalar $\mathcal{O}(a)$ improvement coefficients within the fit. We derived the full NLO ChPT continuum expression in~\cite{Bali:2021qem}. To this order no LECs enter other than those
that we already have determined above.
In view of the limited statistical accuracy, we do not attempt to parameterize higher lattice artifacts and set priors on the six LEC, keeping them close to the central fit results in tab.~\ref{tab:lecs}.
From this fit with $\chi^{2} / N_{\rm df} \approx 1.09$, we obtain at the physical point
\begin{align}
  a_{\eta}(\mu = \infty) = 0.1564\left(\substack{37\\63}\right)_{\rm stat}(45)_{\rm syst}(8 t_{0}^{\rm ph})^{-3/2} = 0.01676\left(\substack{40\\67}\right)_{\rm stat}\left(48\right)_{\rm syst}\left(65\right)_{t_{0}}\,\mathrm{GeV}^3,\\
  a_{\eta^{\prime}}(\mu = \infty) = 0.308\left(\substack{16\\17}\right)_{\rm stat}(80)_{\rm syst}(8 t_{0}^{\rm ph})^{-3/2} = 0.0330\left(\substack{18\\17}\right)_{\rm stat}\left(80\right)_{\rm syst}\left(16\right)_{t_{0}}\,\mathrm{GeV}^3.
  \label{eq:gluonmefitresults}
\end{align}
The systematic error is taken as the difference between the results of this fit and the NLO prediction using the previously determined LECs alone. The fit and the physical point results are shown in fig.~\ref{fig:gluonme}. Results at lower renormalization scales can be found in tab.~19 of~\cite{Bali:2021qem}.

\begin{figure}
  \includegraphics[width=0.49\linewidth]{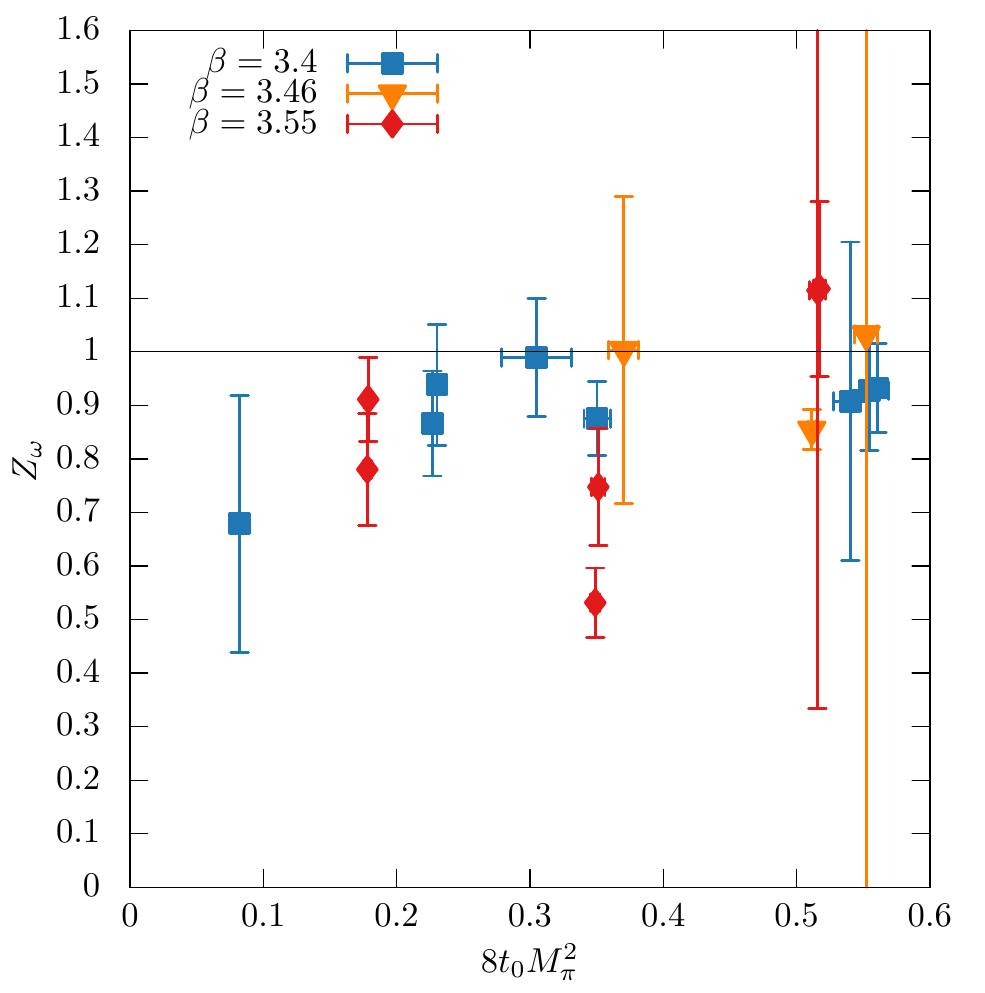}
  \includegraphics[width=0.49\linewidth]{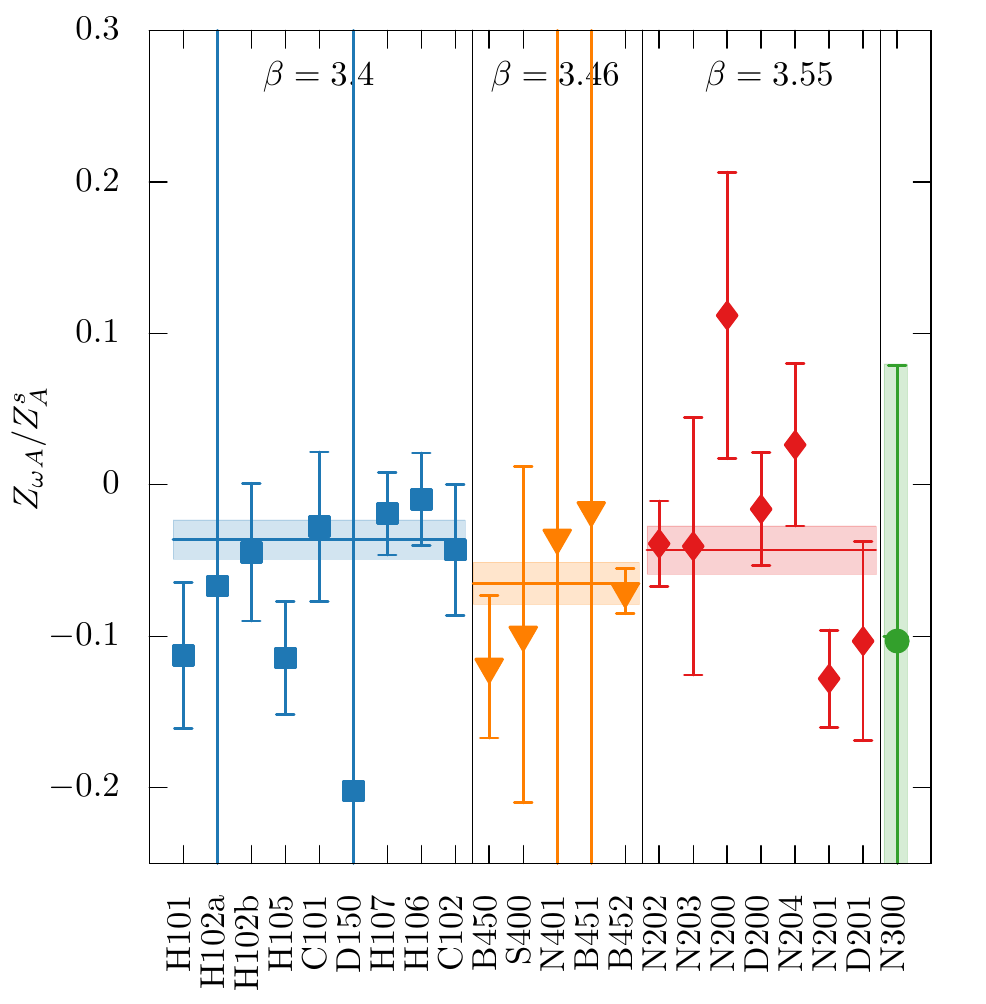}
  \caption{\label{fig:zomegaa} (Left) $Z_\omega$ from solving
    eq.~\eqref{eq:gluonmegluonic}. The $m_s=m_{\ell}$ points are not shown since
    in these cases $F_\eta^0 = 0$ and the equation system is singular.
    (Right) Values of $Z_{\omega A} / Z_A^s$ from solving eq.~\eqref{eq:gluonmegluonic}, assuming $Z_{\omega}=1$.
  }
\end{figure}

\begin{figure}
  \includegraphics[width=\linewidth]{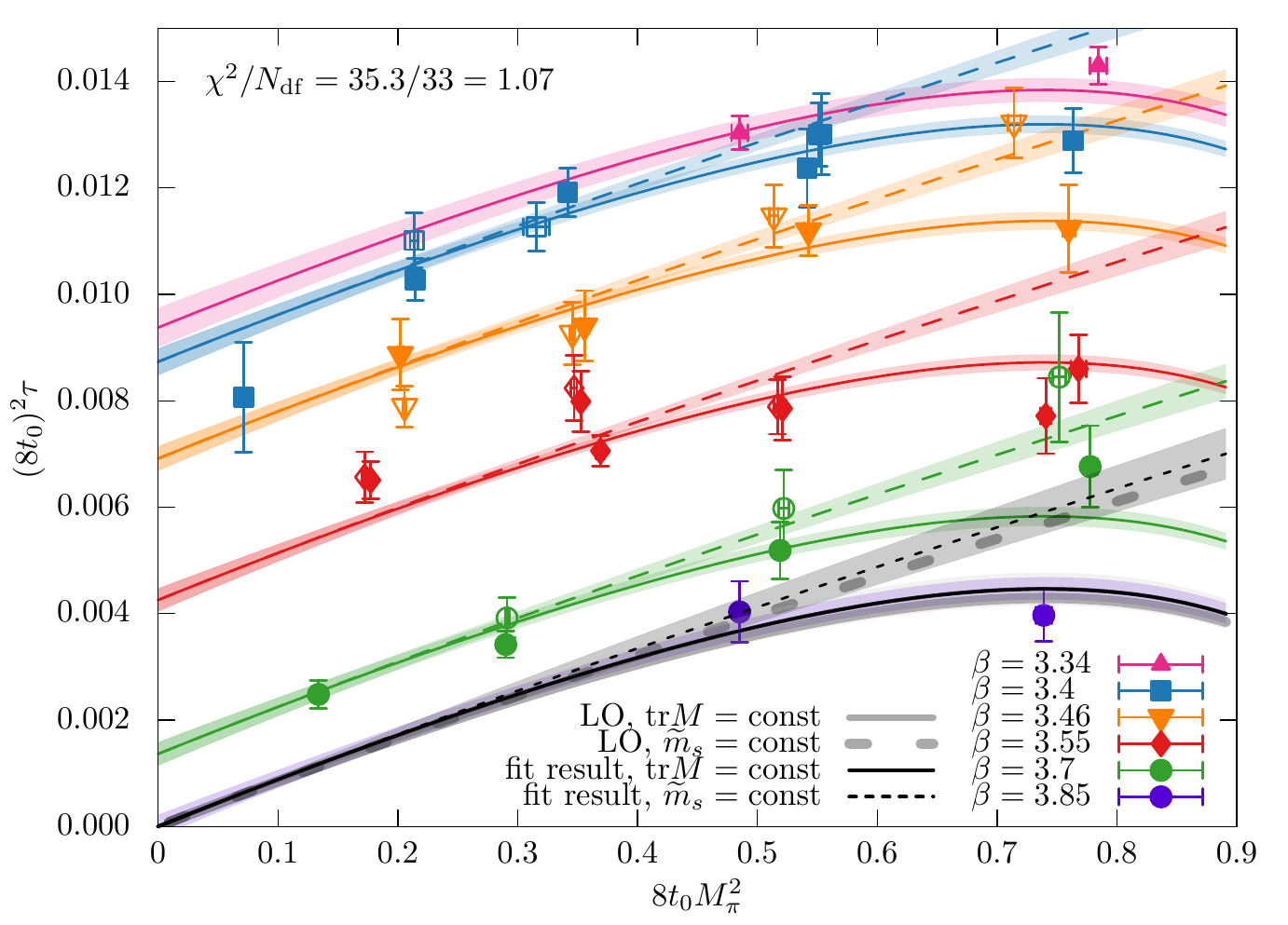}
  \caption{\label{fig:chi}Topological susceptibility for many
    of the CLS ensembles described in~\cite{spectrum}. Filled symbols mark
    ensembles that are simulated with a
    constant sum of quark masses (solid lines), open symbols correspond to
    ensembles with the strange quark mass fixed to approximately the physical
    value (dashed lines).
    Lines and shaded regions are the result of a fit to eq.~\eqref{eq:chifit},
    adding $a^{2}$, $a^{3}$ and $a^{4}$ coefficients.
    The continuum limit result (black lines) is very close to both the fit
    result at
    $\beta=3.85$ as well as the leading order
    expectation (grey lines), when using $\sqrt{8 t_0} F = 0.1866$,
    see~tab.~\ref{tab:lecs}
    and setting $Z_\omega = 1$.}
\end{figure}

To establish the consistency between the above fermionic definition
of the matrix elements $a_{\mathcal{M}}$ and a direct gluonic determination,
we investigate the mixing under renormalization eq.~\eqref{eq:renormomega}.
We start by solving the set of linear equations
\begin{equation}
  a_{\mathcal{M}}(\mu)=2\,Z_{\omega}\langle\Omega|\omega|\mathcal{M}\rangle+
  2\frac{Z_{\omega A}}{Z_A^s}M_{\mathcal{M}}^2F_{\mathcal{M}}^0(\mu),\label{eq:gluonmegluonic}
\end{equation}
for the unknown renormalization factors using the fermionic definition of $a_{\mathcal{M}}$ for $\mathcal{M} = \eta$ and $\mathcal{M} = \eta^{\prime}$.
The result for $Z_{\omega}$ is plotted on the left panel of fig.~\ref{fig:zomegaa} and is consistent with $Z_{\omega} = 1$, which we expect when using flowed gauge fields.
We confirm this observation by fitting the topological susceptibility on many CLS ensembles to the leading order
ChPT continuum expectation for the
$N_{f}=2+1$ theory~\cite{DiVecchia:1980yfw,Leutwyler:1992yt},
\begin{equation}
    \hat{\tau} = \frac{F^2}{2} \left(\frac{1}{2 M_K^2 - M_\pi^2} + \frac{2}{M_\pi^2} \right)^{-1},
    \label{eq:chifit}
\end{equation}
where we parameterize the sizable cutoff effects with quadratic, cubic
and quartic terms in the lattice spacing. Data and fit are
shown in fig.~\ref{fig:chi}. We obtain
$F\sqrt{8 t_{0}^{\chi}} / Z_{\omega} = 0.190(13)$, which agrees with
$F\sqrt{8 t_{0}^{\chi}} = 0.1866(48)$ from the parametrization of the
masses and decay constants (see tab.~\ref{tab:lecs}) if we
set $Z_{\omega} = 1$, which demonstrates that indeed $Z_{\omega}\approx 1$
with the gradient flow definition of $\omega$.

\begin{figure}
  \includegraphics[width=0.49\linewidth]{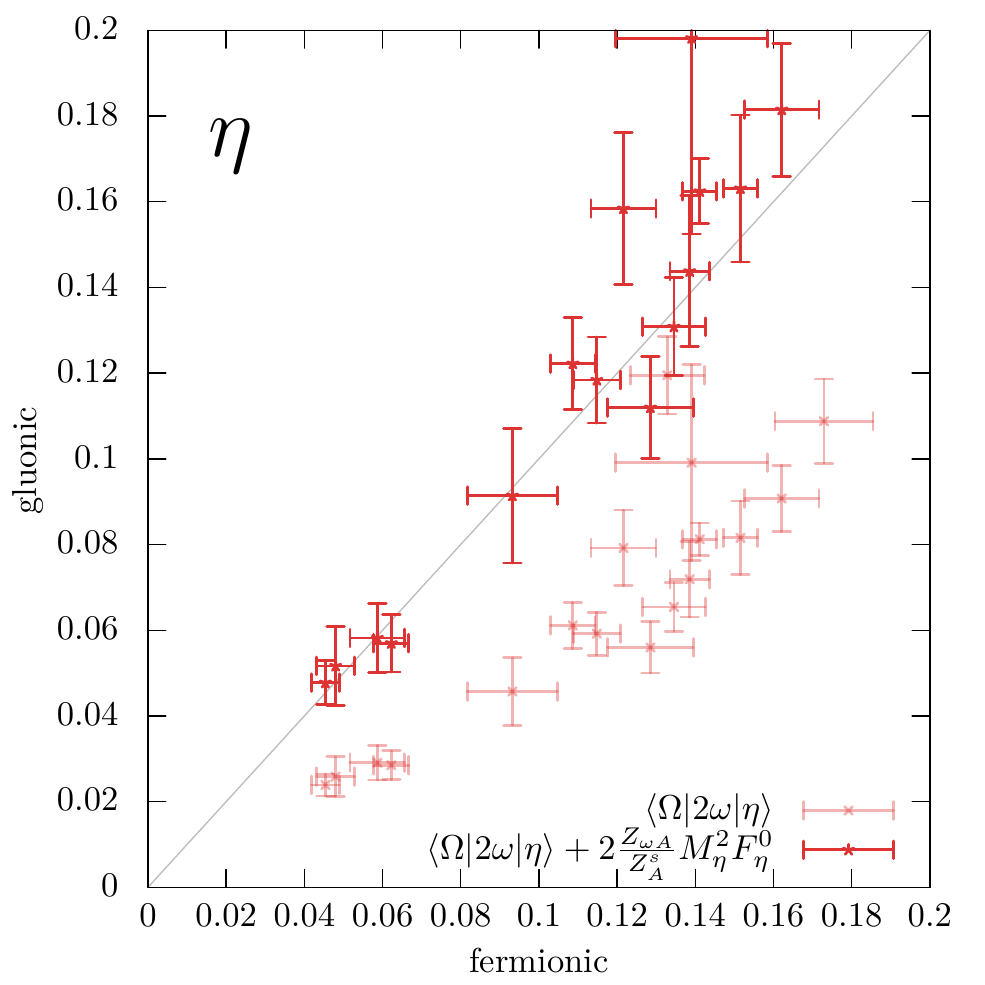}
  \includegraphics[width=0.49\linewidth]{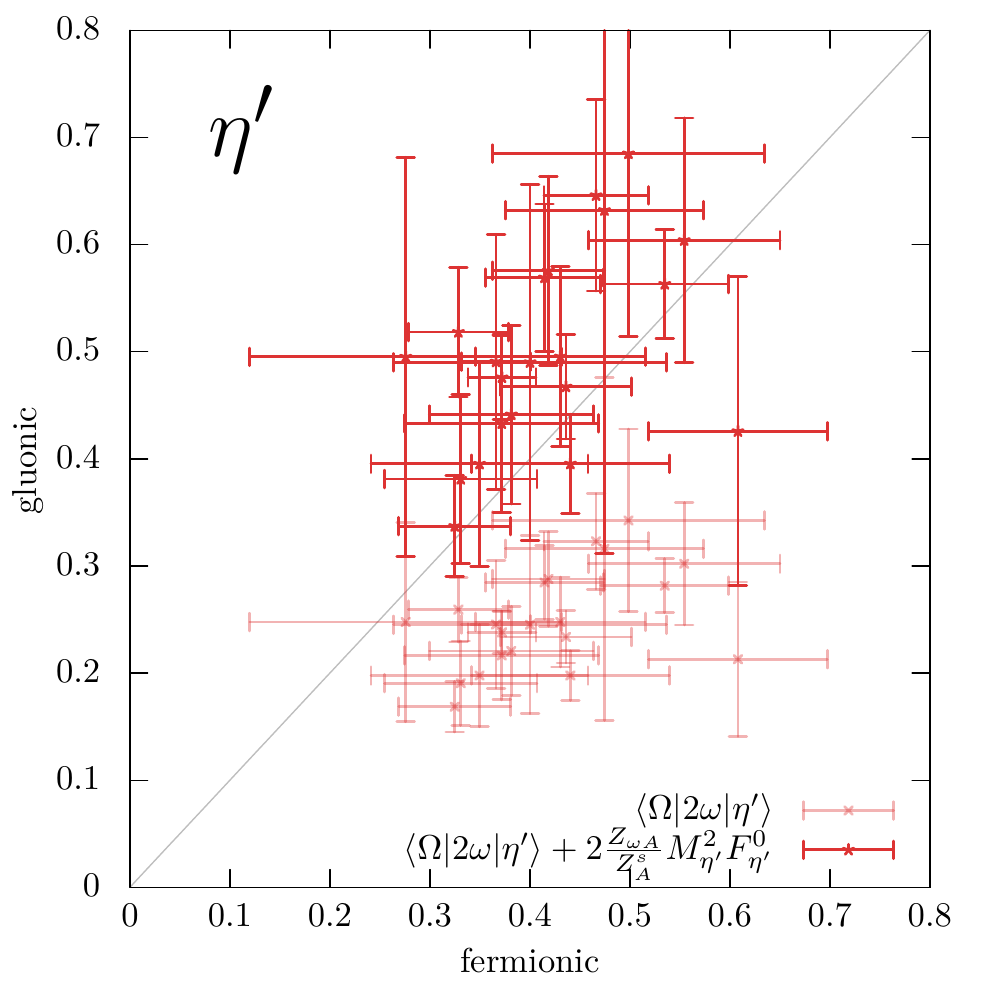}
  \caption{Scatter plot of the fermionic (eq.~\eqref{eq:gluonmefermionic},
    horizontally) and gluonic (eq.~\eqref{eq:gluonmegluonic}, vertically)
    determinations of the gluonic matrix elements $a_{\eta^{(\prime)}}$. The
    mixing with the derivative of the axialvector current is non-negligible:
    the unrenormalized lattice matrix elements (pale red points) do not
    agree with the
  fermionic definition.\label{fig:gluonmecmp}}
\end{figure}

Assuming $Z_{\omega} = 1$ and using the fermionic definition of $a_{\mathcal{M}}$,
we isolate the ratio $Z_{\omega A} / Z_{A}^{s}$ in eq.~\eqref{eq:gluonmegluonic},
that only depends on the coupling parameter but not on the scale.
We plot this in the right panel of fig.~\ref{fig:zomegaa}.
Using weighted averages for each lattice coupling, this determines
the second renormalization factor and enables a comparison between
the fermionic (eq.~\eqref{eq:gluonmefermionic}) and gluonic
(eq.~\eqref{eq:gluonmegluonic}) definitions of the anomalous matrix elements,
see the scatter plot fig.~\ref{fig:gluonmecmp}. We also show
the anomalous matrix elements without taking the mixing into account
($Z_{\omega A}=0$). In this case, one would have underestimated
$a_{\eta^{(\prime)}}$ by about 30\%, using the gluonic definition.

\section{Summary}
In these proceedings we summarized our computation of the masses, the complete set of decay constants and the anomalous matrix elements for the $\eta$ and $\eta^{\prime}$ states that we recently published in~\cite{Bali:2021qem}. The interested reader is referred to that publication for more technical detail and comparison to literature values. The decay constants and anomalous matrix elements were determined for the first time from first principles. Our physical point extrapolation is facilitated by using twenty-one CLS ensembles on two mass trajectories and four lattice spacings. We employed NLO large-$N_{c}$ ChPT to describe the continuum limit mass dependence. The results are well parameterized by this
ansatz, including the anomalous matrix elements and we determined all six NLO LECs. The scale dependence has been studied and shows that the FKS approximation~\cite{Feldmann:1998vh,Feldmann:1998sh,Feldmann:1999uf} works surprisingly well at low energies, which is a consequence of $\Lambda_1(1\,\mathrm{GeV})\approx 0$. In addition to the already published results, we determined masses of higher excitations for $m_{\ell}=m_s$, see fig.~\ref{fig:excstates}. We found a very large mass for the second excited octet state while the first excited octet state has a smaller mass than the first excited singlet state.

{\bf\noindent Acknowledgments.}
This work was supported by
the Deutsche Forschungsgemeinschaft (SFB/TRR-55 and FOR 2926) and
the European Union’s Horizon 2020 research and innovation programme under
the Marie Sk{\l}odowska-Curie grant agreement no.~813942 (ITN EuroPLEx)
and grant agreement no.~824093 (STRONG-2020). We thank
our colleagues in CLS [\url{http://wiki-zeuthen.desy.de/CLS/CLS}].
The authors gratefully acknowledge the Gauss Centre for Supercomputing
(GCS) for providing computing time through the John von Neumann
Institute for Computing (NIC) on
JUQUEEN, JUWELS~\cite{juwels} and in particular
JURECA-Booster~\cite{jureca} at Jülich Supercomputing Centre (JSC).
\bibliographystyle{JHEP}
\setlength{\bibsep}{0pt plus 0.3ex}
\bibliography{bibliography.bib}
\end{document}